\documentclass[]{revtex4}%
\usepackage{amsfonts}
\usepackage{amsmath}
\usepackage{graphicx}
\usepackage{amssymb}
\usepackage{braket}
\usepackage{float}%
\providecommand{\U}[1]{\protect\rule{.1in}{.1in}}
\begin{document}
\title{Quantum cosmology of quadratic $f(R)$ theories with a FRW metric}
\author{V. V\'azquez-B\'aez}
\email{manuel.vazquez@correo.buap.mx}
\affiliation{Benem\'erita Universidad Aut\'onoma de Puebla, Facultad de Ingenier\'{\i}a, 72000 Puebla, M\'exico.}
\author{C. Ram\'{\i}rez}
\email{cramirez@fcfm.buap.mx}
\affiliation{Benem\'erita Universidad Aut\'onoma de Puebla, Facultad de Ciencias
F\'{\i}sico Matem\'aticas, P.O. Box 165, 72000 Puebla, M\'exico.}

\begin{abstract}

We study the quantum cosmology of a quadratic $f(R)$ theory with a FRW metric, via one of its equivalent Horndeski type actions, where the dynamics of the scalar field is induced. The classical equations of motion and the Weeler-deWitt equation, in their exact versions, are solved numerically. From the choice of a free parameter in the action follow two cases, inflation + exit and inflation alone. The numerical solution of the Wheeler-DeWitt equation depends strongly on the boundary conditions, which can be chosen so that the resulting wave function of the universe seems to be normalizable and consistent with hermitian operators.
\end{abstract}

\maketitle

\section{Introduction}

Since its formulation general relativity has been a successful theory, verified in many ways and at any scale. However there are instances where it does not reproduce in a precise way the results of observations, in particular the origin of the universe and the early and present inflationary phases. One way to describe these issues have been given by means of modified gravity $f(R)$ theories. Starobisnky \cite{starobinsky} has proposed $f(R)$ theory as an effective action of gravity obtained by coupling it to quantum matter fields, which explains inflation and the reheating after it and recently it has being used to explain the effects of dark matter and dark energy \cite{odintsov,woodard}, see also \cite{felice,sotiriou,Nojiri:2017ncd,Nojiri:2010wj,Nojiri:2006ri} . Even if these theories appear as effective theories, one appealing feature of them is that they are pure gravity theories, although they are higher order. However, under certain conditions it is possible to give actions of Horndeski type equivalent to $f(R)$ actions \cite{langlois,ohanlon}, which have the advantage of being second order and consistently with the usual inflationary or dark energy scenarios where there are scalar fields coupled to Einstein relativity.

Another scenario where a $f(R)$ theory has been considered is the origin of the universe by a tunneling mechanism from ``nothing'' to the de Sitter phase of Starobinski model \cite{vilenkin}. As noted in this work, at this stage a description of the universe is done in the framework of quantum cosmology, from which were computed, in the WKB approximation, the tunneling probability, and the subsequent curvature fluctuations and the duration of the inflationary phase. Quantum cosmology of $f(R)$ theories has been studied also in \cite{fr,kenmoku,biswas}. One frequent problem related to the wave function of the universe is its interpretation and related to it, its normalizability, which might depend on the initial conditions \cite{biswas}.

In this work we consider the quantum cosmology of $f(R)=R+\alpha R^2$ theory in the form with a scalar auxiliary field, in the FRW metric. The equations of motion in this form are second order with an additional degree of freedom, which can be eliminated, leading back to one third order equation. In the case of the Wheeler-deWitt equation, it is usual to do the Hamiltonian formulation in one of the variants with one additional degree of freedom in order to do the Hamiltonian analysis in the usual Dirac formulation. Here we use the formulation due to O'Hanlon \cite{ohanlon}, with an auxiliary scalar field $\phi$, and give a numerical solution for the WdW equation in its exact form. Considering that the scale factor $a$ is positive, we require that the wave function of the universe vanishes at $a=0$, in order that the conjugate momentum of $a$ is hermitic. For the numerical computation, we consider a compact domain in $a$ and $\phi$. We get that the solution is consistent to zero on all boundaries, corresponding to normalizability. This feature persists in the noncompact limit. In the second section we make a short analysis of the classical solutions and in the third section we show the numerical solution of the WdW equation considering values of the parameter $\alpha$ for which classically the solutions are qualitatively different. In the last section we draw some conclusions.

\section{Lagrangian Analysis}
\label{sec_lagrangiano}

Let us focus on the $f(R)$ model for gravity (see \cite{felice,Nojiri:2017ncd}) without matter which has an action given by $\frac{1}{2\kappa^{2}}\int dt \sqrt{-g} f(R)$, variation with respect of the metric $g_{\mu\nu}$ results in
\begin{equation*}
\delta A=\frac{1}{2\kappa^{2}}\int\sqrt{-g}\left[\frac{1}{2}g^{\mu\nu}F(R)+\frac{\partial f(R)}{\partial R}\frac{\delta R}{\delta g_{\mu\nu}}\right]\delta g_{\mu\nu} d^4x,
\end{equation*}
which leads to the following equations of motion
\begin{equation}
F(R)R_{\mu\nu}-\frac{1}{2}f(R)g_{\mu\nu}-\nabla_\mu\nabla_\nu F(R)+g_{\mu\nu}\Box F(R)=0,
\label{ecuacionesf(R)}
\end{equation}
where $F(R)=f'(R)$.

Recalling that for a FRW geometry the metric tensor is the following
\[
g_{\mu\nu}=diag\left(  -N^{2},\frac{a^{2}}{1-kr^{2}},a^{2}r^{2},a^{2}r^{2}%
sen^{2}\theta\right),
\]
then the scalar curvature is given by
\begin{equation*}
\label{curvatura-cal}R=\frac{6}{a^{2}}\left( \frac{a\ddot{a}}{N^{2}}+%
\frac{\dot{a}^{2}}{N^{2}}-\frac{a\dot{a}\dot{N}}{N^{3}}+k\right),
\end{equation*}
which written in terms of the Hubble factor becomes $R=12H^{2}+6\dot{H}$ once we fixed the gauge field to $N=1$ and set $k=0$.
With the scalar curvature as a function of $H$, as stated above, the field equations (\ref{ecuacionesf(R)}) are
\begin{equation*}
3FH^{2}=\frac{1}{2}\left( FR-f \right)-3H\dot{F},
\end{equation*}
\begin{equation*}
-2F\dot{H}=\ddot{F}-H\dot{F},
\end{equation*}
where $F=F(H(t))$ is regarded as a scalar degree of freedon, the so called ''scalaron'' $\varphi$ \cite{starobinsky}.

This approach presents some calculational advantages, for instance it is evident that treating $H$ and $\varphi$ as the dynamical fields the above equations of motion are second order, choosing $a(t)$ as the dynamical variable leads to third order equations; also the dynamics of the scale factor is simply given by $a=\exp\left[\int H dt \right]$, provided one can solve for $H(t)$. Nevertheless this statements are for one side based on setting $k=0$ and on the other side in the fixing on the gauge $N=1$. Let us examine this two conditions.

For $k\neq0$ the curvature in terms of $H$ reads $R=12H^{2}+6\dot{H}+6k/a^2$. Nevertheless, this loss of generality is not a problem since  the observations favor an almost ($k\approx0$) flat universe \cite{universoplano}. On the other side the choice $N(t)=1$ for the gauge field on the Lagrangian formulation represents no difficulty, however if we want to implement the Hamiltonian formulation of the theory, and thus the quantum formulation, we have to keep the gauge arbitrary. Additionally, for the Hamiltonian formulation of an $f(R)$ theory, we should implement Ostrogradsky's formalism \cite{ostrogradsky} to deal with the higher order derivatives in the Lagrangian, for instance the terms containing derivatives of the lapse function must be integrated out of the action since being a gauge field it can not have dynamics, this in turn produces third order derivatives for the scale factor and in general it is not possible to eliminate all the derivatives of $N$ as can be seen in the following Lagrangian
\begin{equation*}
L=-\frac{24\alpha a\dddot{a}}{\dot{a}}{N^{3}+}\frac{12\alpha a\ddot{a}^{2}%
}{N^{3}}+\alpha\dot{a}^{2}\left(  -\frac{24\ddot{a}}{N^{3}}+\frac{72k}%
{aN}+\frac{36a\dot{N}^{2}}{N^{5}}\right)  +\frac{36\alpha\dot{a}^{4}}{aN^{3}%
}+\frac{36\alpha k^{2}N}{a}-\frac{6a\dot{a}^{2}}{N}+6kaN,
\end{equation*}
which corresponds to the quadratic Starobinsky's model, $f(R)=R+\alpha R^{2}$. The third derivative of $a$ can be seen in the first term on the right, also we can see the presence of a derivative of $N$, which complicates the Hamiltonian formulation.
These complications are avoided in the Starobinsky's model \cite{starobinsky}, with the FRW metric plugged in an O'Hanlon type of action  \cite{ohanlon} and given by
\begin{equation}
A=\frac{1}{2\kappa^2}\int \sqrt{-g}\left[R+\phi(\beta\phi+R)\right] \label{action},
\end{equation}
where $\beta$ is a free parameter. This action resembles the action used in \cite{vilenkin}, where the definition of the scalar curvature is regarded as a constraint and used to eliminate $\ddot{a}$ terms in the action.

A variation with respect to $\phi$, results in $\phi=-\frac{1}{2\beta}R$ which leads to
\[A=\frac{1}{2\kappa^2}\int \sqrt{-g}\left(R-\frac{1}{4\beta}R^2\right).
\]
Thus action (\ref{action}) is completely equivalent to that of the Starobinsky model. Once we enter the FRW metric into this action we are able to partial integrate second order derivative terms and all the derivatives of $N$ unlike in the case mentioned above, thus we get
\begin{eqnarray*}
A=\frac{1}{\kappa}\int dt\left[3(1+\phi)(Nak-N^{-1}a{\dot a}^2)-3N^{-1}a^{2}\dot a\dot\phi+\frac{1}{2}\beta N a^{3}\phi^{2}\right].
\label{actionphi}
\end{eqnarray*}

The resulting equations of motion from this action are
\begin{equation}
\left(  6a\dot{a}^{2}+6ka\right)  \left(  1+\phi\right)  +6a^{2}\dot{a}%
\dot{\phi}+\beta a^{3}\phi^{2}=0,\label{eqNactionphi}%
\end{equation}%
\begin{equation}
\left(  4a\ddot{a}+2\dot{a}^{2}+2k\right)  \left(  1+\phi\right)  +2a^{2}%
\ddot{\phi}+4a\dot{a}\dot{\phi}+\beta a^{2}\phi^{2}=0,\label{eqaactionphi}%
\end{equation}%
\begin{equation}
3a^{2}\ddot{a}+3a\dot{a}^{2}+3ka+\beta a^{3}\phi=0,\label{eqphiactionphi}%
\end{equation}
corresponding to the equation from $N$, $a$ and $\phi$, respectively, here we have fixed the gauge to $N=1$. From eq. (\ref{eqphiactionphi}) we get
\begin{equation}
\phi=-\frac{3}{\beta a^{2}}\left(  a\ddot{a}+\dot{a}^{2}+k\right)
,\label{phi}%
\end{equation}
which substituted in eq. (\ref{eqNactionphi}) gives
\begin{equation}
3k^{2}+\dot{a}^{2}\left(  6a\ddot{a}-9\dot{a}^{2}-2\beta a^{2}-6k\right)
+a^{2}\left(  6\dot{a}\dddot{a}-3\ddot{a}^{2}-2k\beta\right)
=0.\label{eqn_phi}%
\end{equation}
Eq. (\ref{eqaactionphi}) is redundant.

\begin{figure}[h]
\centering
\includegraphics[height=5cm,width=9cm]{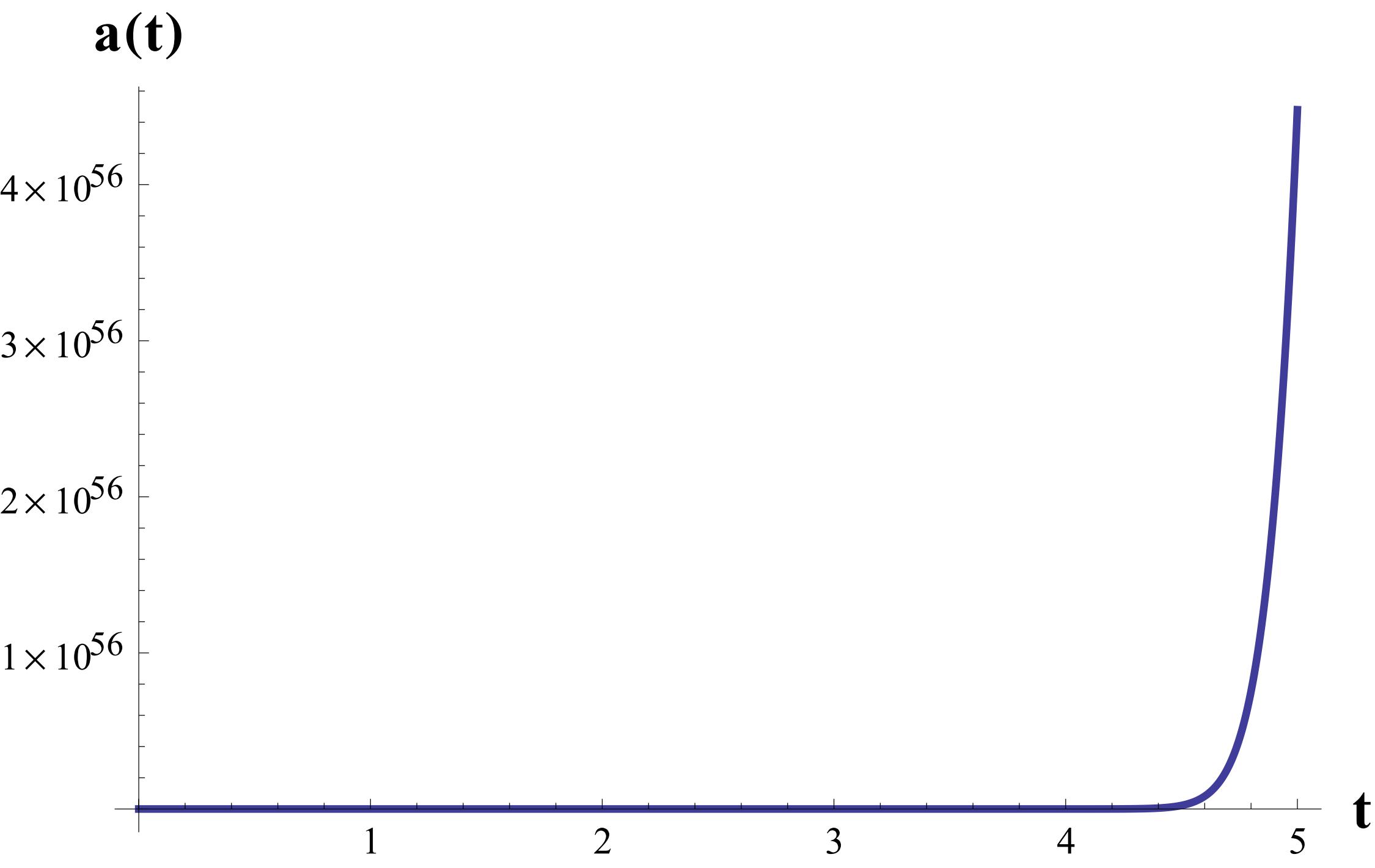}
\caption{{\protect\footnotesize {Numerical profile of $a(t)$ with initial conditions $a(0)=10^{-20}$, $\dot a(0)=10^{-20}$, $\ddot a(0)=10^{-17}$, $\beta=-100$ and $k=0$. }}}%
\label{fig_a_inflacion}
\end{figure}

\begin{figure}[h]
\centering
\includegraphics[height=5cm,width=9cm]{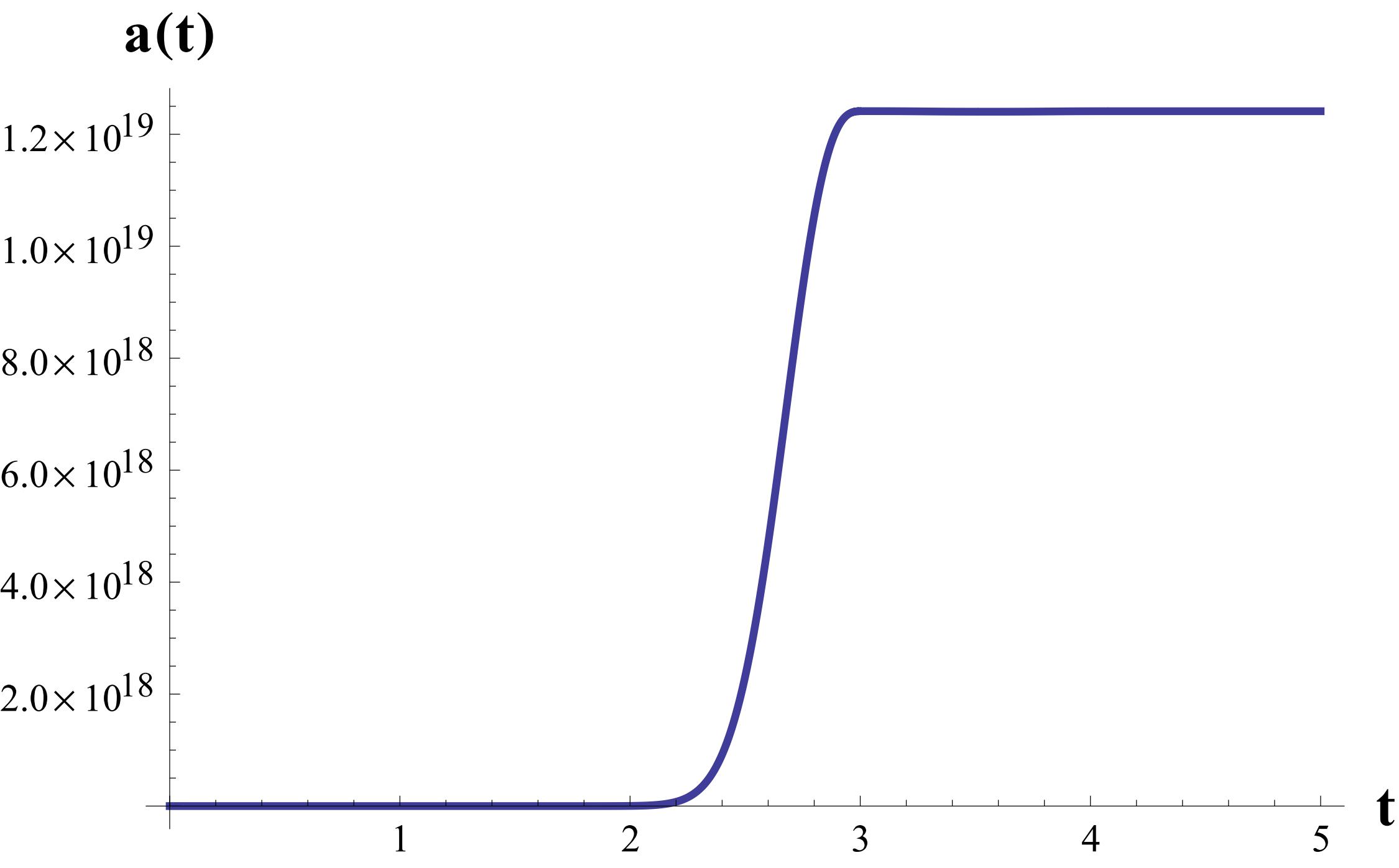}
\caption{{\protect\footnotesize {Numerical profile of $a(t)$ with initial conditions $a(0)=10^{-20}$, $\dot a(0)=10^{-20}$, $\ddot a(0)=10^{-17}$, $\beta=-200$ and $k=0$. }}}%
\label{fig_a_salida}
\end{figure}

Obviously eq. (\ref{eqn_phi}) can be solved only numerically. We have use the liberty to fix the free parameter $\beta$, to our advantage, in order to produce a consistent profile for the evolution of the scale factor. In general, whichever the value of $\beta$, we have an inflation stage as long as $\beta<0$, see figure \ref{fig_a_inflacion}, but for sufficiently large values (in magnitude) of $\beta$ we can see an exit of this inflationary phase, as we depict in figure \ref{fig_a_salida}. Also in this figure we can see that as $a$ reaches certain value its evolution stops, the universe freezes out, this feature of the model should change in the presence of matter, we are analyzing here the very reduced model since the only content of our model is gravity and the curvature scalar ''matter''.

\section{Wheeler-deWitt equation}
\label{sec_hamiltoniano}

In this section we give the quantum formulation of this cosmological model, we begin by computing the canonical momenta of our Lagrangian, they are
\begin{align*}
\pi_{a}  & =-\left[\frac{12a\phi\dot{a}}{N}+\frac{12a\dot{a}}{N}+\frac{6a^{2}\dot{\phi}}{N}\right],\\
\pi_{\phi}  & =-\frac{6a^{2}\dot{a}}{N},\\
\pi_{N} & =0,
\end{align*}
from which, by the usual definition $H(\pi,q)= \pi_{i}q_{i}-L$, we get the customary form of the Hamiltonian as a product of a first class constraint times a Lagrange multiplier, $H=N H_{0}$, with
\begin{equation}
H_{0}=-6 k a \phi -6 k a-\beta  a^3 \phi ^2+\frac{\phi  \pi _{\phi }{}^2}{6 a^3}+\frac{\pi _{\phi }{}^2}{6 a^3}-\frac{\pi _a \pi _{\phi }}{6 a^2}.
\label{hamiltonian}
\end{equation}

Note that besides the advantages mentioned in sec. \ref{sec_lagrangiano} due to the formulation of the model based on $R$ and $\phi$ we have the usual first class constraint $H_{0}=0$ and the lagrange multiplier $N$, there is no presence of additional constraints nor we had the need to use Ostrogradky's method or other schemes of generalized mechanics.

In the quantum version, the action of the first class constraint $H_{0}$ becomes a condition on the state wave function of the system, thus we have $\widehat{H}\psi=0$ which is the Wheeler-DeWitt equation. Therefore, we promote the classical quantities in eq. (\ref{hamiltonian}) to quantum operators in the coordinate representation ($ q\to \widehat{q}, \text{ } \pi\to -i \hbar \partial_{q} $); since the classical Hamiltonian contains terms which represent ambiguities in the quantum scheme we chose the Weyl ordering of the operators. Thus we have the non linear equation
\begin{equation}
\widehat{H}\psi =\left[  -\hbar^{2}\left(\phi+1\right) \frac{\partial^{2}}{\partial\phi^{2}}+a\hbar^{2}\frac{\partial^{2}}{\partial\phi\partial a}-2\hbar^{2}\frac{\partial}{\partial\phi}-6\beta a^{6}\phi^{2}-36 k a^{4}\left(  \phi+1\right)  \right]  \psi=0.
\end{equation}

\begin{figure}[h]
\centering
\includegraphics[height=5cm,width=9cm]{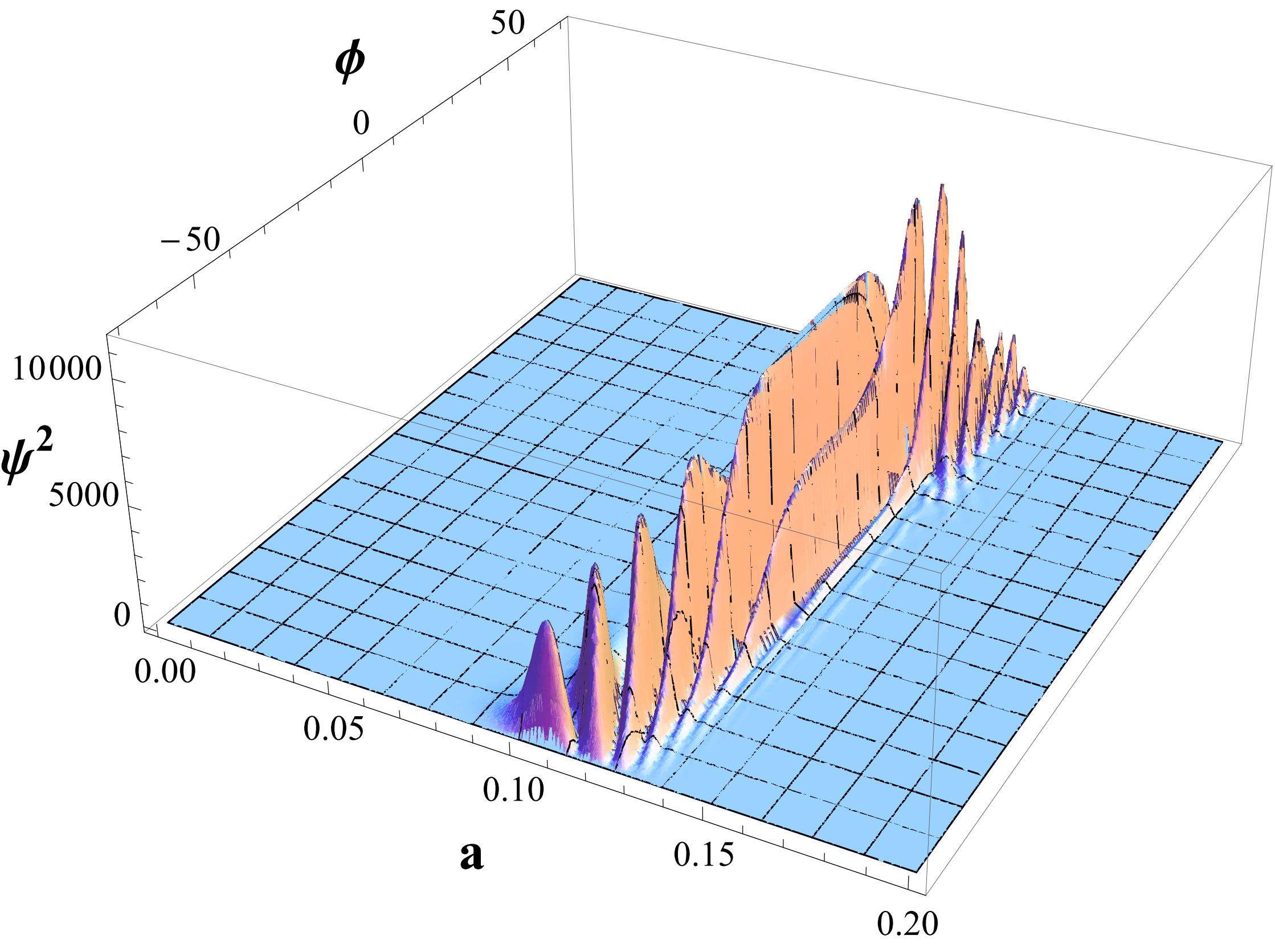}
\caption{{\protect\footnotesize {Numerical profile of $\psi(a,\phi)$ with initial conditions: $\psi(0,\phi)=0$ and $\psi(a,\phi_{0})=f(a)$, where $\phi_{0}$ defines half the with of our solution domain over $\phi$. Here we have set $\hbar=1$, $\beta=-200$ and $k=0$.}}}%
\label{psi_salida}
\end{figure}

\begin{figure}[h]
\centering
\includegraphics[height=5cm,width=9cm]{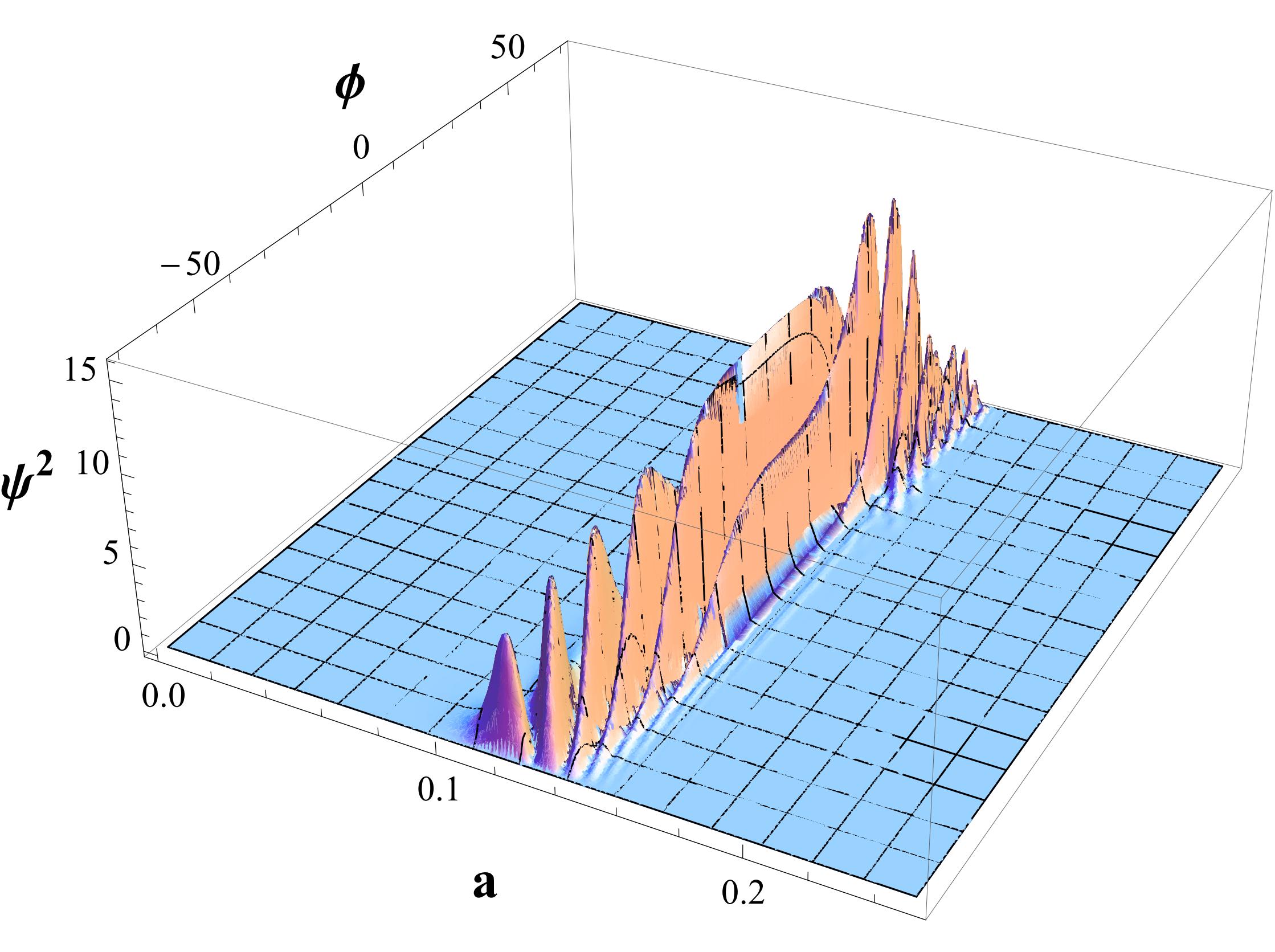}
\caption{{\protect\footnotesize {Numerical profile of $\psi(a,\phi)$ with initial conditions: $\psi(0,\phi)=0$ and $\psi(a,\phi_{0})=f(a)$, where $\phi_{0}$ defines half the with of our solution domain over $\phi$. Here we have set $\hbar=1$, $\beta=-100$ and $k=0$.}}}%
\label{psi_inflacion}
\end{figure}

We solve this numerically and as can be seen from the profile of $\psi^{2}$ depicted in figures (\ref{psi_salida},\ref{psi_inflacion}) the model produces a normalizable wave function of the universe whether we deal with values of $\beta$ producing an exit of the inflationary phase or not, in the classical scheme of the previous section; this means that in principle the model could predict both cases from the quantum version. Note that we have set $\psi(a=0,\phi)=0$ as an initial boundary condition to ensure the hermiticity of the quantum operators as explained in \cite{kuchar} considering that $a\geq 0$. Also we have set $\psi(a,\phi_{0})=f(a)$ on the $\phi$-boundaries, this is a good condition in order to prevent the numerical solution to be identically zero; conditions of the type $\partial_{\phi} \psi(a,\pm \phi_0)=0$ restrict to strong the wave function so that it can not evolve from the zero value on the $a=0$ boundary. Moreover, we can see that the non-vanishing values of $\psi$ are grouped in a well defined region of the domain, which ensures that the wave function is normalizable. Assuming that the universe originates in a very small value of $a$ this wave function points to a fast grow of the universe to the values where this wave function has it's maximum, pointing to the existence of a inflationary phase.

The statement concerning the normalization of the wave function, although motivated by the figures above, can be analytically justified if at least on the upper and lower limits of the numerical domain one could find hints that the boundary conditions we establish hold in certain degree, with the only assumption that the wave function is $C^{2}$-class. We attack this by proposing a series solution to the Wheeler-DeWitt equation given by
\begin{equation}
\psi={\displaystyle\sum\limits_{n=0}^{\infty}}f_{n}(\phi)a^{n},
\end{equation}
once entered in (\ref{hamiltonian}) result in a system of differential equations for the coefficients $f_{n}(\phi)$
\begin{align*}
\left(  1+\phi\right)  f_{0}^{\prime\prime}+2f_{0}^{\prime}  & =0\\
\left(  1+\phi\right)  f_{1}^{\prime\prime}+f_{1}^{\prime}  & =0\\
f_{2}^{\prime\prime}  & =0\\
\left(  1+\phi\right)  f_{3}^{\prime\prime}+f_{3}^{\prime}  & =0\\
\left(  1+\phi\right)  f_{4}^{\prime\prime}-2f_{4}^{\prime}+36k\left(
1+\phi\right)  f_{0}  & =0\\
& \vdots
\end{align*}
which can be analytically solved starting from $f_{0}$ and up to the desired $n$-coefficient, resulting in
\begin{align*}
f_{0}  & =c_{2}-\frac{c_{1}}{\phi+1},\\
f_{1}  & =c_{3}\log(\phi+1)+c_{4},\\
f_{2}  & =c_{5}\phi+c_{4},\\
f_{3}  & =c_{5}\left(  \frac{\phi^{2}}{2}+\phi\right)  +c_{6}\\
f_{4}  & =\frac{18c_{2}k\phi^{2}}{\hbar^{2}}+\frac{36c_{2}k\phi}{\hbar^{2}%
}-\frac{18c_{1}k\phi}{\hbar^{2}}+\frac{c_{6}\phi^{3}}{3}+c_{6}\phi^{2}%
+c_{6}\phi+c_{7}\\
& \vdots
\end{align*}

The free parameters $c_{n}$ can be fixed according to the boundary conditions on the wave function, to ensure that $\psi(0,\phi)=0$ we must have $c_{1}=c_{2}=0$, we also set $c_{3}=0$ in order to prevent the presence of the term $\log(1+\phi)$, the rest of these constants remain unconstraint, we set $c_{n}=(-1)^{n}/n!$, with these one can see that around $a=0$ the wave function converges to zero. In the same way one can calculate the same series solution around $a=\infty$ which results in the trivial solution, all the coefficients are identically zero. This can be easily seen from the corresponding Wheeler-DeWitt equation
\begin{equation*}
\left[ \hbar^{2}b^{6} \left(  1+\phi\right)  \frac{\partial^{2}}{\partial\phi^{2}}+\hbar
^{2}b^{7}\frac{\partial^{2}}{\partial\phi\partial b}+2b^{6}\hbar^{2}%
\frac{\partial}{\partial\phi}+36b^{2}k\left(  1\text{$+$}\phi\right)
\text{$+$}6\beta\phi^{2}\right]\psi=0,
\end{equation*}
where $b=1/a$, which at $b=0$ gives $\psi=0$.

Also we can see from eq. (\ref{hamiltonian}) that as $\phi \to \infty$ the $\phi^{2}$ term dominates, forcing the wave function to be zero in this limit and then allowing the integrability over $\phi$. Thus, we can say that the function is normalizable since being a soft surface it must reach a finite maximum and then take on zero values asymptotically  as $a \to 0 $ and $a \to \infty$.

\section{Conclusions}

We have studied the classical and quantum formulation of a $f(R)$ modified theory of General Relativity  based on the Starobinsky model, with a scalar field in a Horndeski type action, in a cosmological setting with a FRW metric. Thus the lagrangian and hamiltonian formulations are straightforward. We consider the numerical solutions for the exact equations in both scenarios, classical and quantum, taking a compact domain for the numerical computation. Although the dependence of the classical solutions on the parameter $\alpha$ can change their topology, the corresponding solutions of the WdW equation do not show such a difference. In fact, for suitable boundary conditions, these solutions tend to zero at the boundaries, pointing to normalizability of the wave function, consistently with the probabilistic interpretation. The form of the wave function suggests that this case could be interpreted by a conditional probability as in \cite{susyfrw}, where the scalar field plays the role of time.


\end{document}